\documentclass[twocolumn]{aastex631}

\usepackage{xcolor}
\usepackage{refcount}
\usepackage{booktabs}
\usepackage{amsmath}
\usepackage{ulem}

\shorttitle{SS Cyg}
\shortauthors{Dutta et al.}
\graphicspath{{./}{figures/}}

\begin{document}

\title{A broadband X-ray study of the dwarf nova SS Cyg during quiescence and outburst}

\author[0000-0001-9526-7872]{Anirban Dutta}
\affiliation{Astronomy and Astrophysics Group, Raman Research Institute Sadashivnagar, Bangalore, Karnataka-560094}

\author[0000-0003-1703-8796]{Vikram Rana}
\affiliation{Astronomy and Astrophysics Group, Raman Research Institute, Sadashivnagar, Bangalore, Karnataka-560094}

\author[0000-0002-8286-8094]{Koji Mukai}
\affiliation{CRESST II and X-ray Astrophysics Laboratory, NASA/GSFC, Greenbelt, MD 20771, USA}
\affiliation{Department of Physics, University of Maryland Baltimore County, 1000 Hilltop Circle, Baltimore MD 21250, USA}

\author[0000-0002-6211-7226]{Raimundo Lopes de Oliveira}
\affiliation{Departamento de F\'isica, Universidade Federal de Sergipe, Av. Marechal Rondon, S/N, 49100-000, S\~ao Crist\'ov\~ao, SE, Brazil}
\affiliation{Observat\'orio Nacional, Rua Gal. Jos\'e Cristino 77, 20921-400, Rio~de~Janeiro, RJ, Brazil}








\begin{abstract}

We present a broadband X-ray study ($\sim$\,0.3-50 keV) of the dwarf nova SS Cyg highlighting the changes in the accretion during two phases, the quiescence and the outburst states. The investigation was based on simultaneous observations carried out with the XMM-Newton and NuSTAR telescopes in two epochs, involving medium and high-resolution spectroscopy. Spectra were harder during quiescence ($kT_{\rm high}\sim22.8$ keV) than outburst ($kT_{\rm high}\sim8.4$ keV), while the  mass accretion rate increased by $\sim35$ times in outburst ($1.7\times10^{16}  \rm g\;s^{-1}$) than quiescence. The bolometric luminosity (0.01-100.0 keV) during the outburst was dominated by a blackbody emission ($kT_{\rm BB}\sim28$ eV) from the optically thick boundary layer, and the inner edge of the accretion disk resides very close to the WD surface. X-rays from the accretion disk boundary layer are consistent with the white dwarf having mass $1.18_{-0.01}^{+0.02} \rm M_{\odot}$. Our study conclusively confirms the presence of the reflection hump in the 10-30 keV range for both phases, which arises when X-ray photons hit colder material and undergo Compton scattering. We estimated a similarly strong reflection amplitude during quiescence ($\sim1.25$) and outburst ($\sim1.31$), indicating both the WD surface and disk are contributing to reflection. The neutral Fe K$_{\alpha}$ line, which is correlated with Compton reflection, also showed similar strength ($\sim80$ eV) in both phases. Finally, X-rays also revealed the presence of a partial intrinsic absorber during the outburst, possibly due to an outflowing accretion disk wind.

\end{abstract}

\keywords{Dwarf novae(418) --- Cataclysmic variable stars(203) --- X-ray binary stars(1811)}

\section{Introduction} \label{sec:intro}

Cataclysmic Variables (CVs) are a class of semi-detached binary systems in which a white dwarf (WD) accretes material from the generally late-type main sequence secondary by Roche-lobe overflow mechanism. 
Among the CVs are the Dwarf Novae (DNe), consisting of `non-magnetic' WD, with magnetic field $\lesssim 10^{5}$ G, and exhibiting frequent optical outbursts \citep{Warner_1999, Hellier_2001}. This is the case of our target, SS Cyg. 

The most popular theory behind the outbursts in DNe is the ``Disk Instability Model" \citep{Osaki_1974, BathandPringle_1982a, BathandPringle_1982b, BathandPringle_1982c, Osaki_1996}. The accretion disk remains calm and tranquil in the normal (quiescence) phase, continuously receiving mass from the secondary, at a rate that is assumed to be constant. However, the mass accretion rate onto the white dwarf is lower, so the gradual accumulation of the material inside the disk increases the surface density, thus temperature. After a critical density is reached, a chain reaction kicks in, increasing temperature, viscosity, ionization, and opacity rapidly, and a heating wave sets in. This thermal-viscous instability increases the mass transport rate through the disk, and ultimately increases the mass accretion rate by the WD. The increased mass accretion onto the WD triggers the higher flux in X-rays along with the increased optical counterpart which generally starts early with the increased mass transfer inside the accretion disk. 

DNe accretion theory suggests the presence of a boundary layer (BL) between the inner edges of the disk and the WD surface. The properties and location of BL and its emission are sensitive to the mass accretion rate. During the quiescence (typical mass accretion rate is $\lesssim10^{16} \; \rm g s^{-1}$), a strong shock is formed in the BL due to the velocity mismatch of the WD surface and the Keplarian velocity of the inner edge of the disk, and Keplarian motion energy is transferred into electromagnetic radiation in the form X-rays. The hot ($\sim10^8$ K) gas in optically thin BL cools down by thermal bremsstrahlung emission, mostly in hard X-rays. However, during enhanced mass accretion rate of outburst ($\gtrsim 10^{16} \; \rm g s^{-1}$), the boundary layer becomes optically thick to its own radiation, approaching a blackbody condition, and radiates mostly in very soft X-rays with temperature $\sim10^5$ K \citep{Pringle_1979, Narayan_1993, Popham_1995}. This has been tested with multiple broadband observations \citep{Ricketts_1979, JonesWatson_1992, Wheatley_2003}. Even though the outburst phase produces mostly soft X-rays, there is clear detection of hard X-rays with temperatures up to a few keVs. One possible explanation is due to the vertical density gradients in the accretion disk where a fraction of optically thin accreting materials emit hard X-rays \citep{Patterson_1985a,Patterson_1985b}. Another line of thought by \cite{Ishida_2009, Wheatley_2005} express that hard X-rays arise from the optically thin corona formed over the optically thick BL in outburst.

As the matter cools down before settling onto the WD surface, the X-ray continuum is characterized by multitemperature components. Moreover, a part of this X-ray continuum  undergoes photoelectric absorption by the WD surface and/or accretion disk and emits neutral iron K$_\alpha$ emission at 6.4 keV. Another part gets reflected by WD surface and/or accretion disk via Compton reflection \citep{GeorgeFabian_1991, MattandPerola_1991}. The reflection hump is produced mainly by the Compton scattering of hard X-ray photons to energies around 10-30 keV. This is because, at lower energies, the photoelectric absorption dominates over Compton scattering, and at higher energies (above a few tens of keV), the Compton scattering takes control. The dependence on photoelectric absorption also features a prominent Fe-K edge at 7.0 keV. Since the reflection is a geometrical effect, it is also dependent on the viewing angle of the reflection site other than the elemental abundance of the reflecting gas. 

SS Cyg, one of brightest CV in the sky and a popular dwarf nova with WD mass $(M_{\rm WD})$ $1.19 \pm 0.02 \rm M_{\odot}$, secondary mass $(M_{\rm K})$ $0.704 \pm 0.002 \rm M_{\odot}$ \citep{Friend_1990} ($M_{\rm WD}=0.81\pm0.19 M_{\odot}$ and $M_{\rm K}=0.55\pm0.13 M_{\odot}$ \citep{Bitner_2007}), binary inclination angle $37\degr \pm 5\degr$ and orbital period $6.603 \; \rm hrs$ \citep{Shafter_1983, Hessman_1984, Bitner_2007, Miller-Jones_2013}, has been subject to multiple studies in X-rays (Ginga and ASCA \citep{Done_1997}, RXTE \citep{McGowan_2004}, ASCA \citep{Baskill_2005} and Suzaku \citep{Ishida_2009}, NICER and NuSTAR \citep{Kimura_2021}). The studies by \cite{Done_1997, Ishida_2009, Kimura_2021} of SS Cyg showed the presence of significant Compton reflection during both states. However, these studies have their own limitations arising from data quality or physical interpretation.

Now, using the simultaneous data from XMM-Newton, having good energy resolution, and NuSTAR, having high sensitivity in hard X-rays and with excellent cross-calibration between these two telescopes, we analysed the broadband spectra of SS Cyg during both the quiescence and outburst phases. We model the multi-temperature continuum with plasma emission components and incorporate the contribution from the iron line complex, the soft X-ray features, absorption, and directly measuring Compton reflection. We aim to address the questions like how the accretion and nature of the X-ray emitting plasma changes between two phases and what are the possible reflection sites in either of the two phases.

In this paper, we have described the observation and data reduction in Section \ref{sec:ObsDataRed} , the analysis and the results in Section \ref{sec:AnalysisResults}, the discussion in Section \ref{sec:discussion}, and finally, the conclusions and summary in the last section.

\section{Observation and Data Reduction} \label{sec:ObsDataRed}

\begin{figure}[hbt!] 
		\centering 
		\includegraphics[width=\columnwidth, trim = {0cm 0cm 0cm 0cm}, clip]{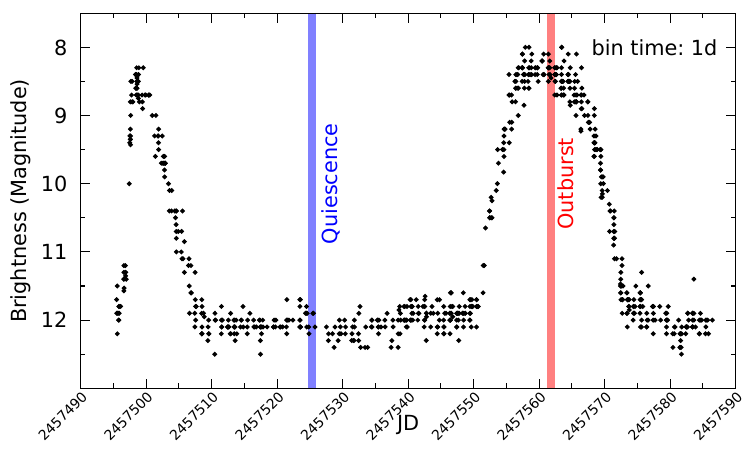}
		\caption{Optical lightcurve of SS Cyg in visual band (Source: \textcolor{blue}{AAVSO}). The epoch of simultaneous NuSTAR and XMM-Newton observations during quiescence and outburst are marked. \label{fig:sscyg_aavso}}
\end{figure}

SS Cyg was observed simultaneously (see Table \ref{tab:Obslog}, PI: Vikram Rana) with XMM-Newton \citep{Jansen_2001} and NuSTAR telescopes \citep{Harrison_2013} in both of its quiescence and outburst phases. Fig. \ref{fig:sscyg_aavso} shows the X-ray observational period of XMM-Newton and NuSTAR during quiescence and outburst along with an optical lightcurve from the American Association of Variable Star Observers (AAVSO). The observation was made to secure the broadband spectra extending from 0.3 keV to 79.0 keV of the X-ray energy band. The good energy resolution of XMM-Newton EPIC (European Photon Imaging Camera) detectors \citep{Struder_2001, Turner_2001} in the low energy band (0.3-10.0 keV), employed with high sensitivity of NuSTAR-FPM detectors in high energy band (staring from 3.0 keV and extending up to 79.0 keV), offer an excellent quality data for analysis. The high-resolution reflection grating spectrometer, RGS \citep{den_Herder_2001}, onboard XMM, provides spectra in the 0.35-2.5 keV energy band, resolving the prominent emission lines produced in the source.

\begin{deluxetable*}{ccccc}
\tablecaption{Observation log of SS Cyg \label{tab:Obslog}}
\tablewidth{0pt}
\tablehead{
\colhead{State} & \colhead{Telescope} & \colhead{Obs ID} & \colhead{Obs Date \& Time} &
\colhead{Exposure (s)}
}
\decimalcolnumbers
\startdata
Quiescence & XMM-Newton & 0791000201 & 2016-05-16 \& 07:18:26 & 30000 \\
 & NuSTAR & 80202036002 & 2016-05-16 \& 06:26:08 & 49789 \\
Outburst & XMM-Newton &  0791000101 & 2016-06-21 \& 18:08:10 & 36000 \\
 & NuSTAR & 80202036004 & 2016-06-21 \& 17:11:08 & 52161
\enddata

\end{deluxetable*}

\subsection{NuSTAR}

The two focusing imaging telescope modules of NuSTAR, FPMA and FPMB,  capable of focusing hard X-rays (3.0-79.0 keV) with very high sensitivity, observed the source. These two telescopes of Wolter-I configuration employ grazing incidence mirrors to bring the X-rays to their focus. Each of the telescope modules is equipped with one detector module, which comprises 4 CZT-detectors. We have selected a 50 arcsec radius circular region to include the source. The 100 arcsec radius circular background region (in the same CZT detector as that of the source) has been selected to extract the background subtracted source data. There are no other X-ray emitting sources present in that detector. The NuSTAR data is devoid of any pile-up issues, thanks to its use of CZT detectors. We have used NuSTARDAS v2.1.2 along with NuSTAR calibration files v20221229 for data calibration, screening and science-products extraction. The default screening criteria for attitude, dead time, orbit and instrument parameters are used to obtain science data products, like the spectra and detector response files. The spectra from each telescope module have been rebinned using \texttt{grppha}  to a minimum of 25 counts per bin to utilize $\chi^{2}$ minimization for spectral fitting. For spectral analysis involving NuSTAR data, both the FPMA and FPMB spectra are used to improve the signal-to-noise ratio.

\subsection{XMM-Newton}

The XMM-Newton satellite consists of three X-ray telescopes. The EPIC-PN detector is placed behind one telescope, receiving X-ray photons with PN-CCD arrays. The EPIC-MOS detectors are two in number and utilize other telescopes for receiving the X-ray photons using Metal Oxide Semiconductor CCD arrays. These three EPIC detectors (one PN, two MOS) have been used to observe the source in 0.3-10.0 keV energy band. The two telescopes, working for MOS1 and MOS2, are also equipped with two grating spectrometers, RGS1 and RGS2. Nearly half of the X-ray flux received  by those two telescopes are sent to MOS detectors whereas, other half is used by the RGS. We have used the first order spectra from the grating spectroscope, which can resolve emission and absorption lines in 0.35-2.5 keV band (5-38 \AA), comprising the important $K_\alpha$ transition from ionised C, N, O, Ne, Mg, Si elements as well as  Fe L-Shell emission lines.

We have used XMMSAS \citep{Gabriel_2004} software (version 20.0.0) for data reduction and extracting science products like the spectra and the detector response files. For the latest calibration files, we have used XMMSAS ccf repository \footnote{\url{https://www.cosmos.esa.int/web/xmm-newton/current-calibration-files}}. The observation was taken in Small Window mode for both PN and MOS detectors. Since SS Cyg is a fairly bright source, we have checked for pile-up in all three detectors using the epatplot tool. We have found that the outburst phase data suffers from pile-up, most strongly in PN. The pile-up effect has been reduced by selecting annular regions to include the source. Since the central core pixels contain maximum counts, thus having maximum impact on pile-up, we have excised those pixels to eliminate the pile-up issue. Running some trials on the annular region size, we have finalized on (15,30) arcsec radii annular region for source selection (for PN, MOS1, MOS2), which minimizes the pile-up without sacrificing the number of counts too much. The background is selected from a circular region in the same CCD as that source, with a radius of 40 arcsec (for PN, MOS1, MOS2). However, since RGS detectors get only ~50\% of the received flux, the pile-up is not noticeable in the RGS data of the outburst phase. This has been cross-checked by comparing the RGS 1st order and 2nd-order spectra. We have checked the quiescence phase data from EPIC and RGS detectors for pile-up, and we didn't notice a significant pile-up there. However, to have a conservative approach, we performed the core-excising technique for the quiescence phase data from EPIC (PN, MOS1, MOS2) detectors as well -- by choosing an annular region of (15,30) arcsec radii for the source and selected a circular background region of 40 arcsec radius in the same CCD as that of the source. The pile-up correction is done following the SAS analysis thread (for EPIC \footnote{\url{https://www.cosmos.esa.int/web/xmm-newton/sas-thread-epatplot}}  and for RGS \footnote{\url{https://www.cosmos.esa.int/web/xmm-newton/sas-thread-pile-up-in-the-rgs}}).

We have also checked for any high background flares present in the data, occurring due to unpredictable solar soft protons. Fortunately, our data does not include any such flaring backgrounds. After employing the necessary selection criteria, we screened and reduced the final science products eg. the spectra and the detector response files. We have applied corrections to the effective area to remove residuals between simultaneous fits of EPIC and NuSTAR observations by applying the command \texttt{applyabsfluxcorr=yes} with the tool \texttt{arfgen}. The spectral data from all the detectors have been re-binned using \texttt{specgroup} to minimum counts of 25  and the oversample parameter of 3.

\section{Data Analysis and Results} \label{sec:AnalysisResults}

In this section, we describe the spectral analysis of the data, performed using XSPEC v12.12.1 \citep{Arnaud_1996}. First, we discuss the analysis of the data obtained during the quiescence phase from NuSTAR and XMM-Newton separately and proceed to simultaneous broadband fitting. Next, we discuss the same for the observations during the outburst. Here, we want to mention that while both XMM observations were carried out entirely with the time interval covered by the respective NuSTAR observations, the latter have data gaps and cover longer periods.  We have also analyzed strictly simultaneous portion of the data and confirm that this did not unduly influence our results. This appears to be because the random varaibility within these observing periods largely averaged out. The only minor exception is in the cross-normalization factor for NuSTAR data for the quiescence observation: the flux was somewhat higher during the first half (with XMM overlap) than during the second half (without). Even so, we do not observe any statistically significant changes in the best-fit model parameters.

The spectral models used in this work are part of the XSPEC: \texttt{mkcflow} \citep[after][]{Mushotzky_1988}, a multi-temperature model assuming cooling flow from a hot optically thin plasma to account for the primary X-ray emission arising from the boundary layer; \texttt{reflect} \citep{Magdziarz_1995}, describing Compton hump expected by reflection; \texttt{tabs} \citep{Wilms_2000}, responding for the photoelectric absorption features of the ISM and local absorbers, eventually associated to the \texttt{partcov} model to partial coverage; \texttt{bbody} to consider the soft X-ray emission from optically thick boundary layer during outburst; \texttt{gaussian} to describe the iron fluorescence line at 6.4 keV; and \texttt{constant} to accommodate and evaluate cross-correlation issues between XMM-Newton and NuSTAR cameras 
We have set the abundance table as Wilms \citep{Wilms_2000} and chosen the Verner photo-electric absorption cross-section \citep{Verner_1996}. As for the \texttt{mkcflow} model, the switch parameter was set to 2 to follow the AtomDB database and the redshift parameter was fixed at $2.67\times10^{-8}$ -- corresponding to the GAIA DR2 distance to SS Cyg adopted in this work, of $114.62$ pc \citep{GAIA_2018}.\footnote{We are aware of the geometric distance determined by \citet{2021AJ....161..147B} from GAIA data of 112.35$^{+0.39}_{-0.35}$ pc, which is only 2\% less than the adopted distance and as such has no impact on the results presented here.} The model flux has been calculated using \texttt{cflux} component from the broadband models. We have extended the range of energy response for all spectral groups in the range 0.01 keV-100.0 keV using \texttt{energies extend} command.

\subsection{Quiescence Phase}
\label{sec:QuiPha}

First, we have looked into the NuSTAR spectra to get an idea of the hard X-ray properties of SS Cyg in quiescence. The data are used in the 3.0-50.0 keV band since photon counts are very low above 50.0 keV and dominated by background. The multi-temperature emission from hot optically thin plasma of the boundary layer has been modelled using cooling flow component \texttt{mkcflow}.  The lower temperature of this component is fixed at the lowest value allowed, 0.0808 keV, since it is closer to the expected lower temperature of the boundary layer plasma while not constrained from the data.  The ISM absorption is incorporated using \texttt{tbabs}. Since the NuSTAR data cannot constrain the column density of the ISM, which influences the spectra below a few hundred of eV, we have fixed its value at $3.5\times10^{19} \: cm^{-2}$ \citep{Mauche_1988}. We have kept the abundance of the cooling flow model free, to reproduce the necessary strength of Fe H-like and He-like $K_{\alpha}$ emission lines. We introduce a narrow \texttt{gaussian} ($\sigma=0$ eV) for the neutral Fe K$_{\alpha}$ line. The resultant best-fit ($\chi^{2}/DOF=719/657$) returns a upper temperature of $\sim33$ keV. However, excess residual for the Compton reflection hump in the hard X-rays ($\sim 10-30$ keV) is evident, and we convolve the \texttt{reflect} component with the plasma emission component to consider Compton reflection. Since the possible reflection sites are regions of white dwarf surface which has freshly accreted material and/or the accretion disk surface, we can assume the same elemental abundance between the Compton reflection component and the plasma emission component. The viewing angle ($i$) of the reflection site is fixed at the default value ($\mu=cosi$\,=\,0.45, which is the ensemble average of the viewing angles of the accreting binary systems) of the reflection component. The final best-fit model ($\chi^{2}/DOF=666/657$) gives upper temperature $21.3_{-1.3}^{+1.5}$ keV, which is in agreement with the finding of \cite{Done_1997} ($21.0_{-5.7}^{+11.0}$ keV) and \cite{Ishida_2009} ($20.4_{-2.6}^{+4.0}$ keV). The abundance comes out to be $0.53_{-0.08}^{+0.09}$ w.r.t solar. The reflection amplitude is obtained as $1.18_{-0.25}^{+0.26}$, validating the presence of reflection as noticed from the excess residual. However, this component may have degeneracy with a complex intrinsic absorber, which can affect even beyond 10 keV for a strong absorber (Fig. 8 of \cite{Dutta_2022}), thereby, the measured value of reflection amplitude can change. Now, in order to probe the effect of the intrinsic absorber in the soft X-rays, we next perform the analysis of the XMM-Newton EPIC spectra.


For the XMM-Newton EPIC data (PN, MOS1, and MOS2), initially, we probed the spectra in 5-9 keV which include the Fe line complex in 6-7 keV. We have used an absorbed bremsstrahlung component to model the continuum and three Gaussian components for the neutral, H-like and He-like Fe K$_{\alpha}$ lines. The column density of the absorber is fixed at the literature value. The fit resulted in a fit-statistic of $\chi^{2}/DOF=110/91$, and the best-fit parameters are quoted in Table \ref{tab:iron}. A spectral plot comparing three iron lines in quiescence and outburst phases are shown in Fig. \ref{fig:iron}. We noticed that the $\sigma$ of all three lines are consistent with EPIC instrument resolution ($\Delta E\sim120$ eV at $\sim6$ keV), and the line widths could not be constrained, so fixed at zero. The line central energies appear at the theoretically expected values, within statistical uncertainty, and the strength of the neutral line is weaker ($82_{-30}^{+34}$eV) compared to the H-like ($102_{-31}^{+62}$eV) and He-like ($131_{-28}^{+46}$eV).

We next model the EPIC spectra in 0.3-10.0 keV energy range. The spectra are described with the cooling flow component (\texttt{mkcflow}) and an absorption component \texttt{tbabs}). The lower temperature is fixed at 0.0808 keV, whereas the upper temperature of the cooling flow component is kept free. We added one narrow Gaussian component for the neutral Fe line. The resultant fit statistics is $\chi^{2}/DOF=530/417$. However, the column density is not constrained, and we obtained an upper limit of $5\times10^{19}\;cm^{-2}$. At this point, to probe the presence of the intrinsic absorber, we included a partial covering photo-electric absorption model (\texttt{partcov*tbabs}), and obtained an improved fit-statistic of $\chi^{2}/DOF=511/415$. The covering fraction of the intrinsic absorber is $0.21_{-0.07}^{+0.09}$ with a column density of $4.2_{-1.5}^{+2.7}\times10^{23}\;cm^{-2}$ indicating that the intrinsic absorber is quite strong. The overall absorber column density, which includes the ISM absorption along the line of sight, is now constrained ( $8.8_{-4.3}^{+5.1}\times10^{19}\;cm^{-2}$). The upper temperature and the elemental abundance of the cooling flow component are found to be $30.7_{-4.0}^{+3.8}$ keV and $0.61_{-0.15}^{+0.16}$ as that of solar values, respectively. 

Now, to check the effect of the reflection and degeneracy with the partial absorber, we introduced the Compton reflection component instead of the partial covering absorber. We obtained a similar fit statistic ($\chi^{2}/DOF=514/416$), with a reflection amplitude of $1.07_{-0.43}^{+0.32}$. The parameter values of the other components are similar (within statistical uncertainty) to that of partial absorber fit. This points to the delicate degeneracy between the partial absorber and the reflection present in the modelling of the EPIC spectra.

\begin{figure}
\centering
\includegraphics[width=\columnwidth, trim = {0.5cm 0.35cm 3.6cm 2.75cm}, clip]{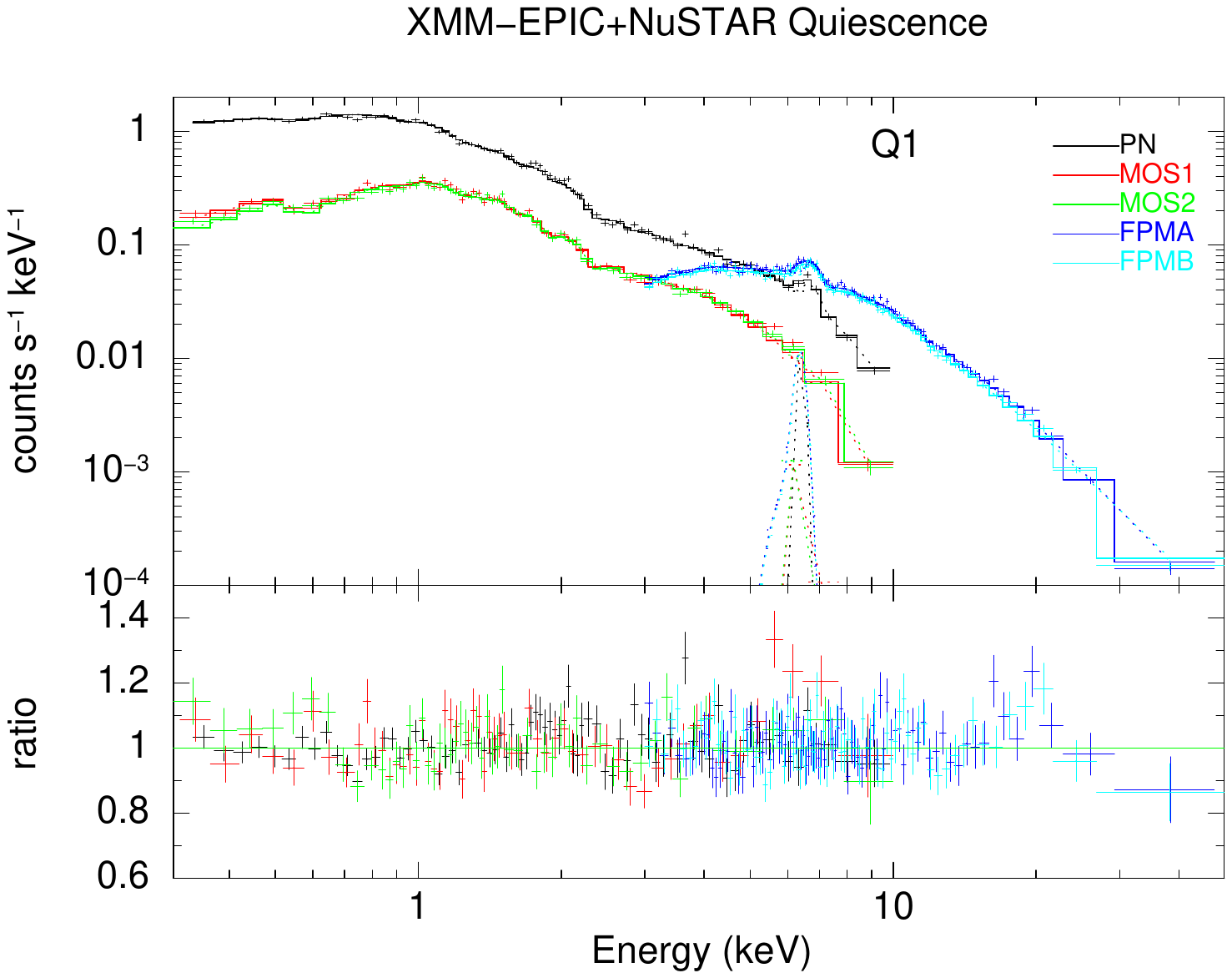}
\includegraphics[width=\columnwidth, trim = {0.5cm 0.35cm 3.6cm 2.75cm}, clip]{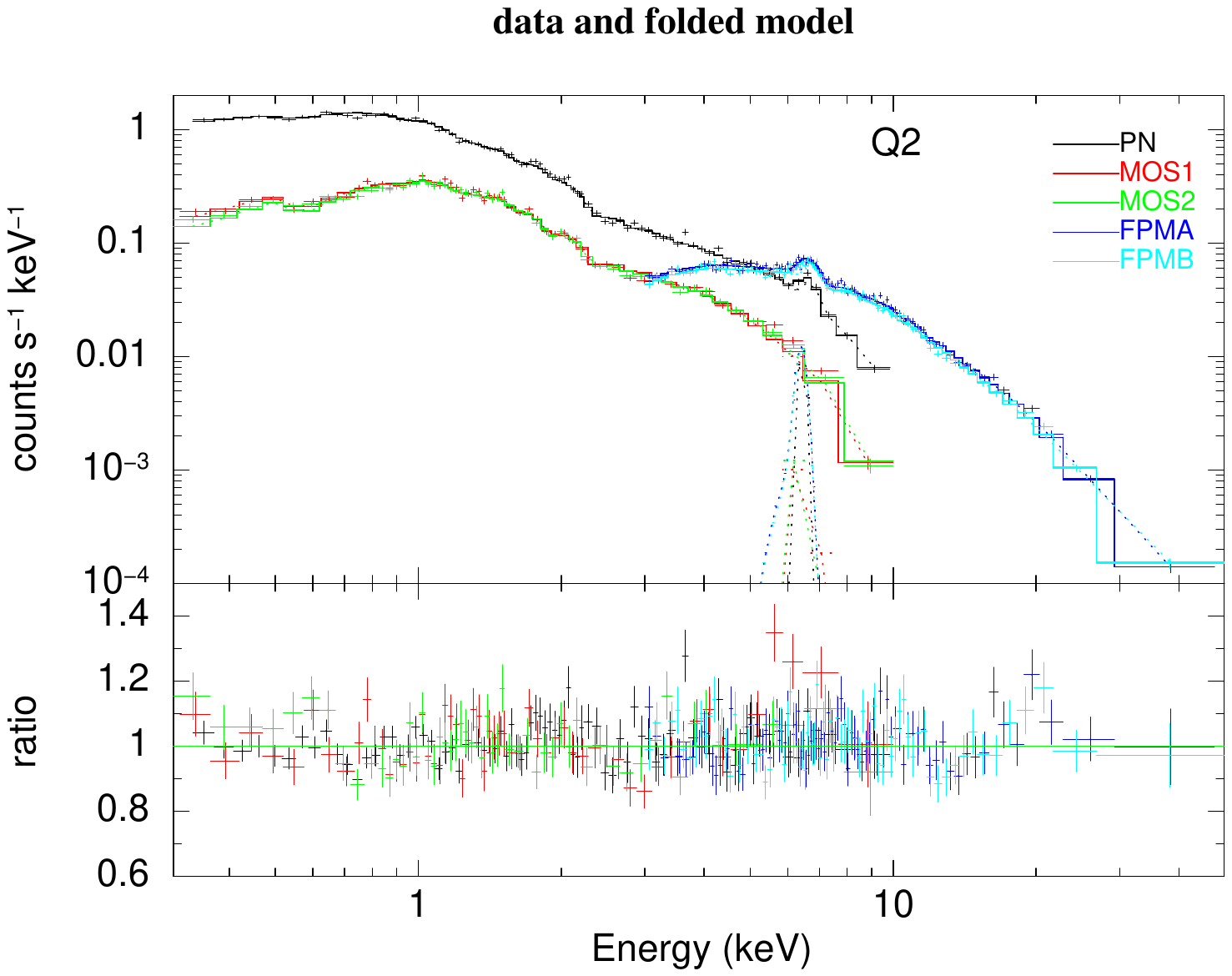}
\caption{Spectrum and ratio (data/model) plot of the quiescence phase joint XMM-EPIC and NuSTAR data including partial covering absorber and Compton reflection, model Q1 (\textcolor{blue}{top}) and including only the Compton reflection component, model Q2 (\textcolor{blue}{bottom})}
    \label{fig:quixmnu}
\end{figure}

In the final stage, guided by the individual modelling of the NuSTAR and EPIC spectra, we proceed towards jointly fitting the simultaneous spectra. Considering the possibility of an intrinsic absorber, we used two model variants - one including the partial absorber and another without it. In XSPEC notations the models are (Q1) \texttt{constant* tbabs* (partcov* tbabs)* (reflect* mkcflow+ gaussian)} and (Q2) \texttt{constant* tbabs* (reflect* mkcflow+ gaussian)}. The best-fit parameter values are quoted in Table \ref{tab:quixmnu} and the spectra are plotted in Fig. \ref{fig:quixmnu}. In the first model, Q1, both the partial absorber and the Compton reflection are considered, where we obtain a reflection amplitude $0.40_{-0.19}^{+0.47}$ and intrinsic absorber column density of $2.6_{-0.6}^{+1.2}\times10^{23}\; cm^{-2}$ with covering fraction of $0.16_{-0.05}^{+0.05}$. For the second model, Q2, the reflection amplitude comes out to be $1.25_{-0.20}^{+0.27}$, which is close to the value obtained from individual EPIC or NuSTAR analysis when intrinsic absorber was not considered. All other parameter values agree within statistical confidence between the models Q1 and Q2. Also, the cross normalisation parameters for model Q2 are: C$_{\rm MOS1}$=$1.02_{-0.01}^{+0.01}$, C$_{\rm MOS2}$=$1.05_{-0.01}^{+0.01}$, C$_{\rm FPMA}$=$1.13_{-0.02}^{+0.02}$, C$_{\rm FPMB}$=$1.09_{-0.02}^{+0.02}$, while C$_{\rm PN}$ is fixed to 1.

At this point, we argue that, since statistically, both these models are equally acceptable, we need to favour one model over another based on physical considerations. This has been discussed in the Section \ref{sec:dis_reflecction}.

\subsection{Outburst Phase}
\label{sec:OutPha}


The outburst NuSTAR spectra in the 3.0-40.0 keV range are fitted with the cooling flow model \texttt{mkcflow} and the ISM absorption model \texttt{tbabs}. 
We restrict the analysis to energies up to 40 keV, after which the background begins to dominate over source counts. We kept the lower temperature of the cooling flow model fixed at 0.0808 keV (for the same reasons explained in Section \ref{sec:QuiPha}) and the column density of the ISM absorber at the literature value, not constrained from the data. We included a \texttt{gaussian} component with free sigma to consider the iron line complex in 6-7 keV, which appears to be a broad peak and could not be resolved by NuSTAR. The fit statistic is $\chi^{2}/DOF=668/554$ with an upper temperature of $10.9_{-0.2}^{+0.2}$ keV, which is expected for the softer spectra in the outburst phase. For the Compton reflection hump, we next included the \texttt{reflect} component with its abundance linked to the cooling flow component, and viewing angle fixed at default value as mentioned in the quiescence phase analysis. We now obtain an improved fit statistic of $\chi^{2}/DOF=650/553$, with reflection amplitude of $0.74_{-0.31}^{+0.39}$. The upper temperature is now  $9.3_{-0.6}^{+0.6}$ keV, and the abundance is $0.43_{-0.11}^{+0.09}$ w.r.t solar values. 

Next, we attempted fitting the iron line complex in 6-7 keV using the XMM-Newton EPIC spectra in the 5.0-9.0 keV range, as done for quiescence phase analysis. The best-fit parameters are quoted in Table \ref{tab:iron}. We notice that  all line widths are consistent with the EPIC instrument resolution, and therefore $\sigma$ was fixed at zero. The strength of the He-like Fe K$_{\alpha}$ line is maximum ($263_{-36}^{+43}$eV) among the three dominant lines. The neutral Fe K$_{\alpha}$ line has similar strength as that during the quiescence phase, whereas the He-like line is much stronger than quiescence. We notice the apparent blueshift ($6.46_{-0.03}^{+0.03}$ keV) of the neutral line. The reason behind is that the K shell lines from the lower ionisation states - Fe XVII through Fe XXII (which are certainly present, as the corresponding L-shell emissions are strongly detected in the RGS spectra of outburst phase, Fig. \ref{fig:rgs}) - have intrinsic centroid around 6.5 keV, which are blended with the neutral line.

We then proceeded to model the entire EPIC spectra (PN, MOS1 and MOS2) in 0.3-10.0 keV. We introduced the cooling flow model to consider the emission from the multi-temperature boundary layer, along with the ISM absorption model. The lower temperature is fixed at 0.0808 keV and the ISM column density is at the literature value. The fit statistic is extremely poor with $\chi^{2}/DOF=6871/438$. We notice that outburst EPIC spectra are significantly complex with an evident signature of blackbody emission, which presents itself as excess in residual in soft X-ray below 0.5 keV. We added the blackbody component, which improved the fit statistic to $\chi^{2}/DOF=4574/437$. There are multiple strong emission line features as expected during the outburst phase of a dwarf nova, like SS Cyg \citep{Mauche_2004, Okada_2008}. Though many lines can be resolved in RGS spectra (see Fig \ref{fig:rgs}), EPIC only detect those as narrow or broad peaks in the soft X-rays below and around $\sim 2$ keV. Therefore, we needed to carefully add multiple emissions components to bring down the fit statistic to an acceptable value. We introduced five Gaussian components to consider the line emission features and obtained an improved fit statistic of $\chi^{2}/DOF=604/426$. Three narrow Gaussian components consider the emission features due to the O-VII line, Si lines and Mg lines. One broad Gaussian accounts for a strong O-VIII line and few neighbouring Fe L-shell emissions, and another broad Gaussian incorporates Ne-lines. At this stage, we include the partial covering component (\texttt{partcov*tbabs}) to probe the presence of the complex intrinsic absorber. The improved fit statistic is now $\chi^{2}/DOF=581/421$, with a column density of the intrinsic absorber $3.2\times10^{22}\;cm^{-2}$ and covering fraction of $0.14_{-0.04}^{+0.04}$.

\begin{figure}
\centering
\includegraphics[width=\columnwidth, trim = {0.5cm 0.35cm 3.6cm 2.38cm}, clip]{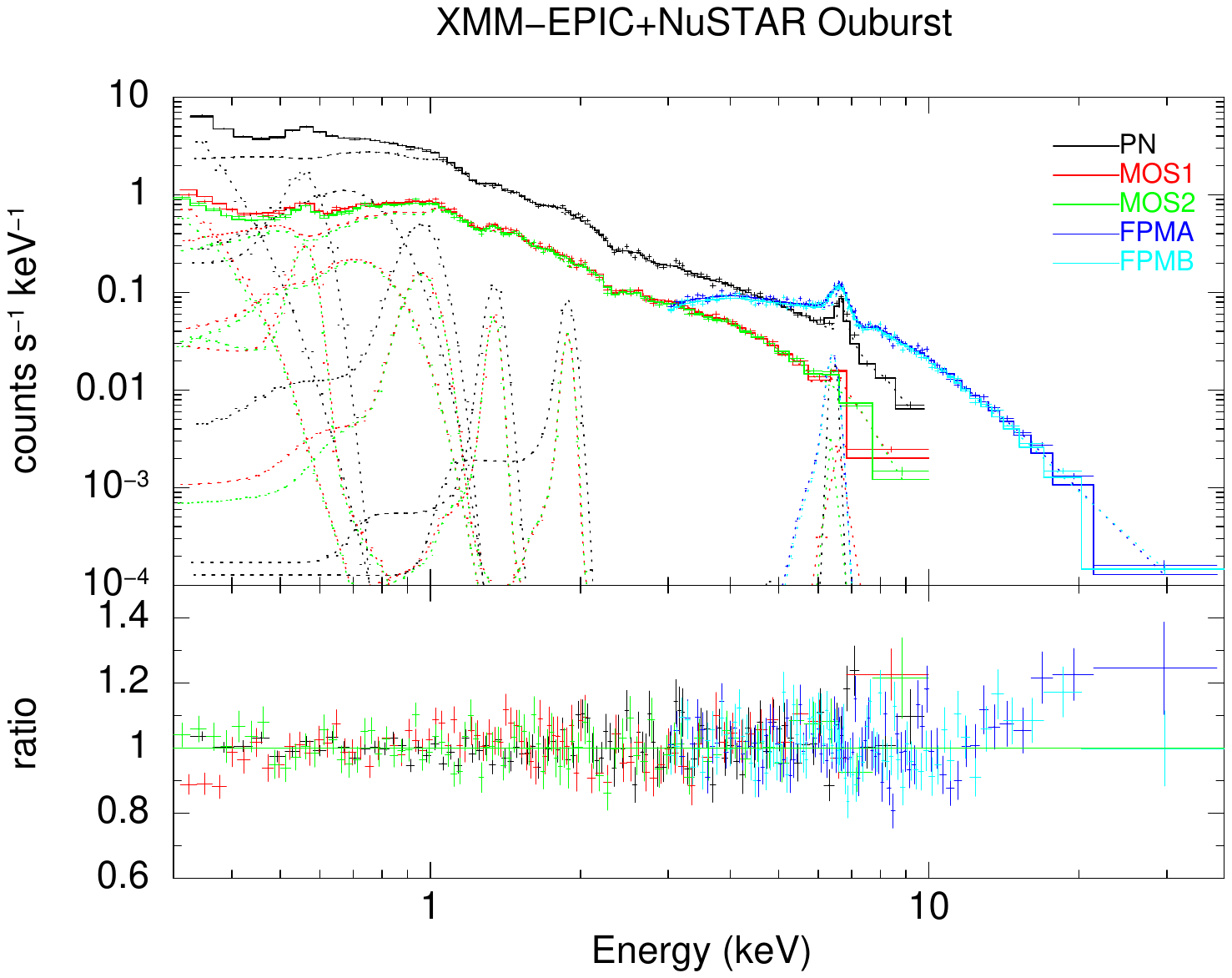}
\caption{Spectrum and ratio (data/model) plot of the outburst phase joint XMM-EPIC and NuSTAR data}
    \label{fig:outxmnu}
\end{figure}

\begin{figure}
\centering
\includegraphics[width=\columnwidth, trim = {0.5cm 0.3cm 3.6cm 2.5cm}, clip]{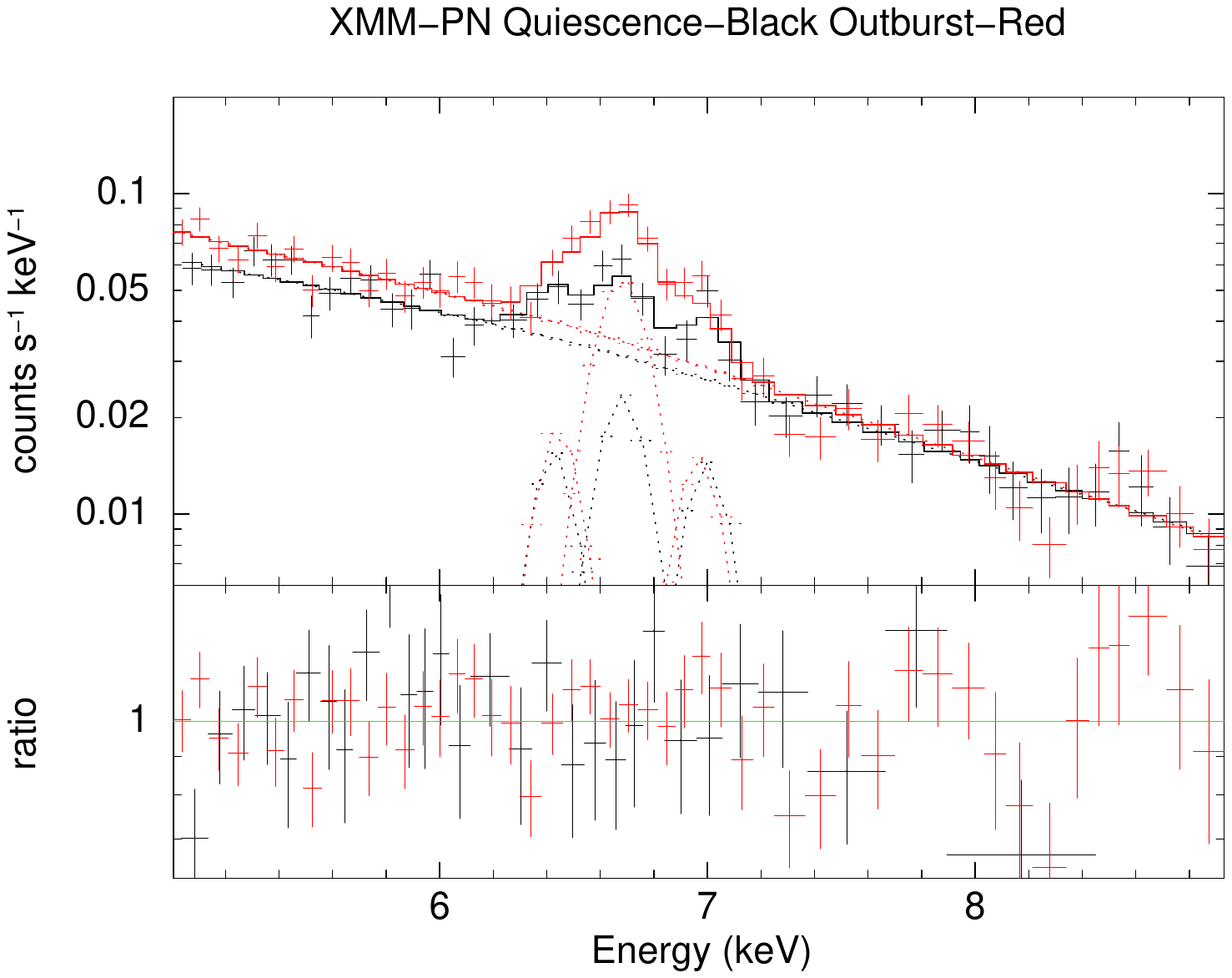}
\caption{Comparison of three iron lines in quiescence and outburst phase. The spectrum and ratio (data/model) plot is shown for PN data only. Quiescence phase data is shown with black, and outburst phase data is shown in red.}
    \label{fig:iron}
\end{figure}

After receiving the clues regarding model components from analysing individual EPIC and NuSTAR spectra, we attempted simultaneous broadband spectral modelling. In XSPEC notation, our model is \texttt{constant* tbabs* (partcov* tbabs)* (reflect* mkcflow+ gauss+ gauss+ gauss+ gauss+ gauss+ gauss+ bbody)}. The fit statistic comes out to be $\chi^{2}/DOF=1257/979$ and best-fit parameter values are quoted in Table \ref{tab:outxmnu}. The spectra are plotted in Fig. \ref{fig:outxmnu} The fit statistic is acceptable for the purpose of our study, where we focus on obtaining the overall description of the continuum. The cross normalisation parameters are: C$_{\rm MOS1}$=$1.04_{-0.01}^{+0.01}$, C$_{\rm MOS2}$=$1.11_{-0.01}^{+0.01}$, C$_{\rm FPMA}$=$1.21_{-0.02}^{+0.02}$, C$_{\rm FPMB}$=$1.18_{-0.01}^{+0.02}$, while C$_{\rm PN}$ is fixed to 1. In this broadband modelling, we considered similar five Gaussian components, as that of EPIC only fit, to consider the excess residual due to line emission features in soft X-rays. The strong blackbody emission is also modelled in the broadband spectra. The reflection amplitude is constrained at $1.31_{-0.30}^{+0.30}$. The abundance ($0.51_{-0.02}^{+0.03}$) is statistically consistent with the corresponding value of that parameter during quiescence. We have also checked that fitting the broadband quiescence and outburst data together with a tied abundance value recovers, to within statistical uncertainties, the results from individual broadband fits, in terms of all the parameters including the abundance ($0.46^{+0.06}_{-0.01}$) and the reflection amplitudes ($1.18^{+0.13}_{-0.12}$ in quiescence and $1.49^{+0.17}_{-0.24}$ in outburst). We could not constrain the overall absorber column density, whose fit value was reaching the lowest limit allowed for that component; so was fixed at the ISM column density from the literature. The intrinsic absorber is evidently required in our fitting, which if not considered, produces a much worse fit statistic of $\chi^{2}/DOF=1358/980$ along with an unphysically high value of reflection amplitude ($\sim2.5$). The upper temperature of the cooling flow component is $\sim 8.4$ keV, less than the quiescence phase (temperature $\sim 22.8$ keV), indicating softer spectra during the outburst.

\subsection{High resolution RGS spectra during quiescence and outburst Phase}

\begin{figure}
\centering
\includegraphics[width=\columnwidth, trim = {0.5cm 0.35cm 3.6cm 2.5cm}, clip]{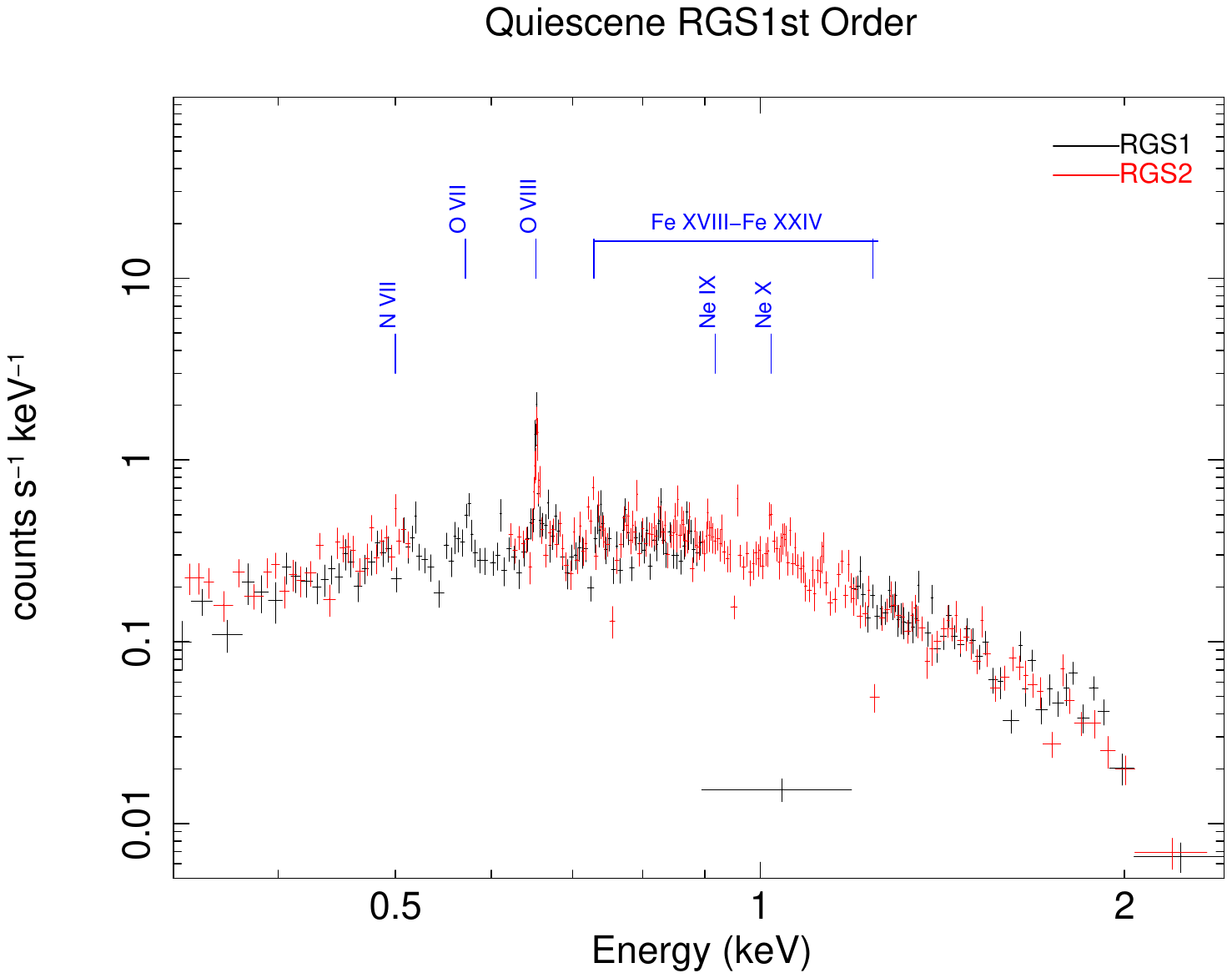}
\includegraphics[width=\columnwidth, trim = {0.5cm 0.35cm 3.6cm 2.5cm}, clip]{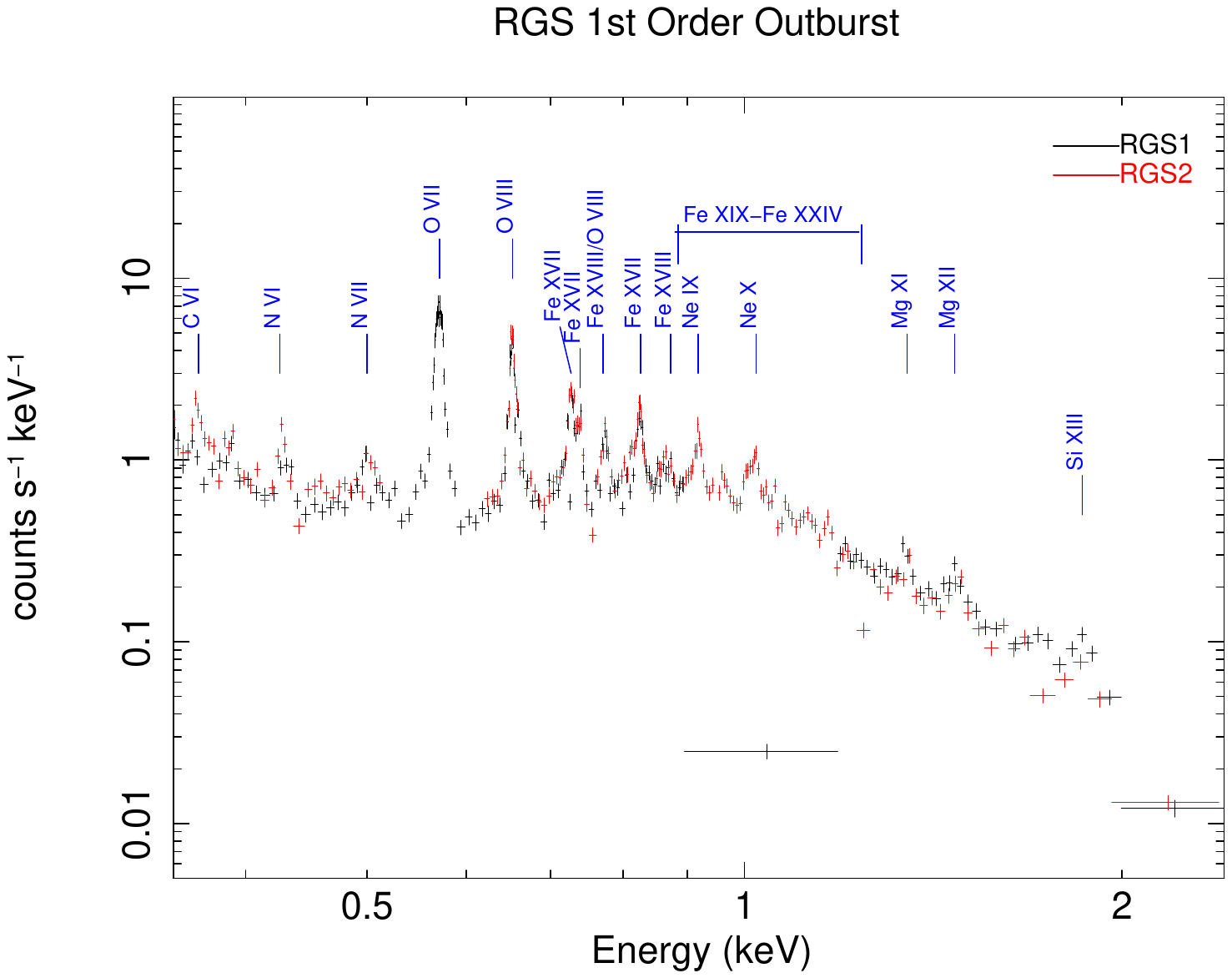}
\caption{Multiple strong line emission features seen in the RGS spectra during quiescence(\textcolor{blue}{top}) and outburst (\textcolor{blue}{bottom}) phase.}
    \label{fig:rgs}
\end{figure}

To verify the line emission features that appeared in the soft X-rays obtained from EPIC, we further checked the RGS spectra. The RGS spectra in 0.35-2.5 keV range for both phases are shown in Fig. \ref{fig:rgs}. We notice that during the quiescence phase, the O-VIII line appears to be the strongest with other weak emission lines. During the outburst, multiple strong emission lines are present, including K$_{\alpha}$ lines of highly ionised states of several low Z elements, e.g., C, N, O, Ne, Mg, Si, and Fe L shell emissions. However, given the complexity involved in modelling the line emissions from the dwarf nova systems, particularly in the outburst phase, we did not attempt robust modelling of the RGS spectra, which is out of scope for this work.

\begin{table}
\centering
\caption{Best-fit parameter values from quiescence joint XMM-EPIC and NuSTAR fit (0.3-50.0 keV) \label{tab:quixmnu}}
\begin{tabular}{cccc}
\hline
\hline
Parameters                         & Unit               & Q$_1^{\hspace{0.1cm} \dagger c}$ & Q$_2^{\hspace{0.1cm} \dagger d}$ \\ \hline
nH$_{\rm tb}$                      & $10^{19}$cm$^{-2}$ & $ 12.5_{-2.2}^{+4.0} $               & $ 15.8_{-2.7}^{+2.8} $               \\
nH$_{\rm pcf}$                     & $10^{22}$cm$^{-2}$ & $ 25.7_{-5.8}^{+11.9} $              &                                      \\
pcf                                &                    & $ 0.16_{-0.05}^{+0.05} $             &                                      \\
T$_{\rm low}$                      & keV                & $ 0.0808_f $                         & $ 0.0808_f $                         \\
T$_{\rm high}$                     & keV                & $ 25.9_{-2.3}^{+1.1} $               & $ 22.8_{-1.1}^{+1.5} $               \\
N$_{C}^{\hspace{0.1cm} \dagger a}$ & $10^{-11}$         & $ 1.46_{-0.06}^{+0.13} $             & $ 1.32_{-0.04}^{+0.05} $             \\
Z                        & Z$_{\odot}$                   & $ 0.56_{-0.13}^{+0.04} $             & $ 0.44_{-0.04}^{+0.06} $             \\
rel$_{\rm refl}$                   &                    & $ 0.40_{-0.19}^{+0.47} $             & $ 1.25_{-0.20}^{+0.27} $             \\
E$_{L}$                            & keV                & $ 6.40_{-0.02}^{+0.04} $             & $ 6.42_{-0.04}^{+0.02} $             \\
$\sigma$                           & eV                 & $ 0_f $                              & $0_f$                                \\
N$_{L}^{\hspace{0.1cm} \dagger b}$ &                    & $ 2.5_{-0.4}^{+0.4} $                & $ 2.4_{-0.4}^{+0.4} $                \\ \hline
$\chi^{2}/DOF$                    &                    & $ 1189/1076 $                      & $ 1202/1078 $   \\ \hline                     
\end{tabular}
\tablecomments{$\dagger a$ : Norm of MCKFLOW (in $M_{\odot}\;yr^{-1}$) \\
    $\dagger b$ : Norm of Gaussian (in total $photons\,cm^{-2}\,s^{-1}$)\\
    $\dagger c$ : Best-fit model using partial absorber and Compton reflection.\\
    $\dagger d$ : Best-fit model using Compton reflection. The model description is in the text.
    }
\end{table}

\begin{table}
\centering
\caption{Best-fit parameter values from joint outburst XMM-EPIC and NuSTAR fit (0.3-40.0 keV) \label{tab:outxmnu}}
\begin{tabular}{ccc}
\hline
Parameters       & Unit               & Values                      \\ \hline
nH$_{\rm tb}$    & $10^{19}$cm$^{-2}$ & $3.5_f$                     \\ \hline
nH$_{\rm pcf}$   & $10^{22}$cm$^{-2}$ & $ 3.4_{-0.9}^{+1.0} $       \\
$\rm pcf$     &                    & $ 0.23_{-0.02}^{+0.02} $    \\ \hline
rel$_{\rm refl}$ &                    & $ 1.31_{-0.30}^{+0.30} $    \\ \hline
T$_{\rm BB}$     & eV                 & $27.9_{-1.7}^{+1.2}$        \\
N$_{BB}^{\hspace{0.1cm} \dagger a}$         & $10^{-3}$          & $24.2_{-7.2}^{+18.1}$       \\ \hline
T$_{\rm low}$    & keV                & $ 0.0808_f $                \\
T$_{\rm high}$   & keV                & $ 8.4_{-0.3}^{+0.3} $       \\
N$_{C}^{\hspace{0.1cm} \dagger b}$          & $10^{-11}$         & $ 5.2_{-0.1}^{+0.1} $       \\
Z                & Z$_{\odot}$              & $ 0.51_{-0.02}^{+0.03} $    \\ \hline
E$_1$            & keV                & $ 0.563_{-0.002}^{+0.006} $ \\
$\sigma_1$       & eV                 & $ 0_f $                     \\
N$_{1}^{\hspace{0.1cm} \dagger c}$          & $10^{-3}$          & $ 2.1_{-0.2}^{+0.1}$        \\ \hline
E$_2$            & keV                & $ 0.647_{-0.029}^{+0.016} $ \\
$\sigma_2$       & eV                 & $ 135_{-19}^{+32} $         \\
N$_{2}^{\hspace{0.1cm} \dagger c}$          & $10^{-3}$          & $ 3.3_{-0.4}^{+1.0} $       \\ \hline
E$_3$            & keV                & $ 1.89_{-0.01}^{+0.01} $    \\
$\sigma_3$       & eV                 & $ 0_f $                     \\
N$_{3}^{\hspace{0.1cm} \dagger c}$          & $10^{-3}$          & $ 6.8_{-1.6}^{+1.6} $       \\ \hline
E$_4$            & keV                & $ 1.34_{-0.01}^{+0.01} $    \\
$\sigma_4$       & eV                 & $ 0_f $                     \\
N$_{4}^{\hspace{0.1cm} \dagger c}$          & $10^{-3}$          & $ 7.6_{-1.8}^{+1.8} $       \\ \hline
E$_5$            & keV                & $ 0.958_{-0.016}^{+0.015} $ \\
$\sigma_5$       & eV                 & $ 58_{-14}^{+15} $          \\
N$_{5}^{\hspace{0.1cm} \dagger c}$          & $10^{-3}$          & $ 0.6_{-0.2}^{+0.2} $       \\ \hline
E$_{L}$          & keV                & $ 6.44_{-0.01}^{+0.01} $    \\
$\sigma$         & eV                 & $0_f$                       \\
N$_{L}^{\hspace{0.1cm} \dagger d}$          & $10^{-5}$          & $ 4.2_{-0.4}^{+0.5} $       \\ \hline
$\chi^{2}/DOF$  &                    & $ 1257/979 $               \\ \hline
\end{tabular}
\tablecomments{$\dagger a$ : Norm of blackbody (L$_{39}$/D$_{10}^2$ where L$_{39}$ is in $10^{39}$ erg s\,$^{-1}$ and D$_{10}$ is in 10 kpc \\
    $\dagger b$ : Norm of MCKFLOW (in $M_{\odot}\;yr^{-1}$) \\
    $\dagger c $ : Norm of Gaussian (in $photons\,cm^{-2}\,s^{-1}$) components, added for the considering the soft X-ray emission features.  \\
    $\dagger d$ : Norm of gaussian (in $photons\,cm^{-2}\,s^{-1}$) \\ 
    }
\end{table}

\begin{table}
\centering
\caption{Iron line comparison from outburst and quiescence XMM-EPIC data in 5-9 keV \label{tab:iron}}
\begin{tabular}{ccccc}
\hline
        & Parameters                         & Unit               & Quiescence               & Outburst                 \\ \hline
        & nH$_{\rm tb}$                      & $10^{19}$cm$^{-2}$ & $3.5_f$                  & $3.5_f$                  \\
        & T$_{\rm Brems}$                    & keV                & $ 11.6_{-3.0}^{+5.6} $   & $7.1_{-0.5}^{+1.7}$      \\ \hline
Neutral & E$_{L}$                            & keV                & $ 6.42_{-0.03}^{+0.03} $ & $ 6.46_{-0.03}^{+0.03} $ \\
line    & $\sigma$                           & eV                 & $ 0_f $                  & $0_f$                    \\
        & eqw                                & eV                 & $ 82_{-30}^{+34} $       & $ 81_{-22}^{+23}$        \\
        & N$_{L}^{\hspace{0.1cm} \dagger a}$ & $10^{-5}$          & $ 2.4_{-0.8}^{+0.8} $    & $3.0_{-0.5}^{+1.1}$      \\ \hline
He-like & E$_{L}$                            & keV                & $ 6.68_{-0.02}^{+0.03} $ & $ 6.69_{-0.01}^{+0.01} $ \\
line    & $\sigma$                           & eV                 & $ 0_f $                  & $ 0_f $                  \\
        & eqw                                & eV                 & $ 131_{-28}^{+46} $      & $ 263_{-33}^{+42} $      \\
        & N$_{L}^{\hspace{0.1cm} \dagger a}$ & $10^{-5}$          & $ 3.7_{-0.8}^{+0.9} $    & $ 8.4_{-1.0}^{+1.1} $    \\ \hline
H-like  & E$_{L}$                            & keV                & $ 6.99_{-0.03}^{+0.03} $ & $ 6.93_{-0.02}^{+0.03} $ \\
line    & $\sigma$                           & eV                 & $ 0_f $                  & $ 0_f $                  \\
        & eqw                                & eV                 & $ 102_{-31}^{+62} $      & $ 86_{-28}^{+36} $       \\
        & N$_{L}^{\hspace{0.1cm} \dagger a}$ & $10^{-5}$          & $ 2.5_{-0.9}^{+0.8} $    & $ 2.8_{-0.7}^{+0.8} $    \\ \hline
        & $\chi^{2}/DOF$                    &                    & $110/91$                & $ 118/97 $              \\ \hline
\end{tabular}
\tablecomments{$\dagger a$ : Norm of gaussian (in $photons\,cm^{-2}\,s^{-1}$) \\ 
    }
\end{table}

\section{Discussion} \label{sec:discussion}

\subsection{Quiescence vs Outburst X-ray : Hard vs Soft} \label{sec:dis_xray}

In dwarf novae systems like SS Cyg, where the material is accreted via an accretion disk, the kinetic energy of the accreting material in the boundary layer is equal to half of the gravitational potential energy of the WD. Following the formation of a strong shock, a fraction of this kinetic energy is transformed into thermal energy, and the remaining kinetic energy gradually converts into thermal energy as the matter cools down, giving rise to the multi-temperature nature of the plasma. If we assume model Q2 during quiescence (the reason discussed in the Sec. \ref{sec:dis_reflecction}), the upper temperature of the cooling flow component ($22.8_{-1.1}^{+1.5}$ keV) is related to the shock temperature of the optically thin plasma in the boundary layer \citep{Pandel_2005}. By relating the thermal energy of the accreting material with the gravitational potential energy of WD  and incorporating the empirical correction factor ($\alpha=T_{\rm high}/T_{\rm s,k}$, where $T_{\rm high}$ is the maximum temperature obtained from plasma emission model, $T_{\rm s,k}$ is the Keplerian shock temperature) of 0.611 \citep{Mukai_2022}; we can write the following equation \citep{Yu_2018, Byckling_2010}:

\begin{equation}
    KT_{high}=\alpha \times \frac{3}{16} \times\frac{G M_{\text{WD}} \mu m_{\text{P}}}{R_{\text{WD}}} \label{eqn:shock}
\end{equation}
where K, G, $\mu$ and m$_{\text{P}}$ represents the Boltzmann constant, gravitational constant, mean molecular weight ( assumed to be 0.615) of the accreting material, and mass of proton, respectively. $M_{\text{WD}}$ and $R_{\text{WD}}$ denote mass and radius of WD respectively. Coupling this equation with the mass-radius relationship for WD \citep{Nauenberg_1972}, we obtain a WD mass of $1.18_{-0.01}^{+0.02}M_{\odot}$ and a WD radius of $0.40_{-0.02}^{+0.01} \times10^9$cm. Our measured mass value agrees very well with the value obtained by \cite{Friend_1990} ($1.19_{-0.02}^{+0.02} M_{\odot}$) using radial velocity study of sodium doublet line. On the other hand, the mass measurement by \cite{Bitner_2007} using absorption line based optical spectroscopic study is smaller ($0.81\pm0.19 M_{\odot}$) than ours. However, we should note that the above two methods rely on the indirect way of measuring the WD mass by performing the difficult task of precise measurement of the inclination angle and mass ratio of SS Cyg. Whereas our X-ray spectroscopy method of WD mass estimate is more direct; by measuring the shock temperature obtained from broadband X-ray spectra which is directly linked to the accretion onto the WD (thereby, linked to its mass). However, we want to caution the reader that the empirical correction factor, which we have used and definitely needed, may not be universal for all DNe \citep{Mukai_2022}.

Our spectral model during the quiescence measures an unabsorbed bolometric flux of $5.9_{-0.1}^{+0.1}\times10^{-11}\;erg\;cm^{-2}\;s^{-1}$ in 0.01-100.0 keV energy range. Considering a distance of $114.62$pc to the source, this flux value corresponds to a luminosity of $\sim9.3\times10^{31} \; \rm erg\;s^{-1}$. Now, relating the luminosity with the mass accretion rate as follows,

\begin{equation}
L=\frac{G M_{\text{WD}} \dot{m}}{2 R_{\text{WD}}} \label{eqn:luminosity}
\end{equation}
we obtain a mass accretion rate ($\dot{m}$) of $\sim4.7\times10^{14} \; \rm g\;s^{-1}$ $\sim 7.5\times10^{-12}\;M_{\odot}\;yr^{-1}$. This value is consistent with the mass accretion rate obtained from the \texttt{mkcflow} normalisation parameter after incorporating the empirical correction factor (actual mass accretion rate =  empirical correction factor $\times$ cooling flow normalization; since cooling flow $T_{\rm high}$ is empirical factor times the actual shock temperature, therefore $T_{\rm high}$ gives an apparently lower $M_{\rm WD}/R_{\rm WD}$ ratio, thus apparently higher mass accretion rate).

Now, for the outburst phase, broadband spectral analysis indicates an upper temperature of the optically thin plasma to be $8.4_{-0.3}^{+0.3}$ keV. This result points out the spectral softness during the outburst than quiescence. The possible origin of this hard X-ray emitting plasma is the optically thin corona instead of the boundary layer, which is optically thick during this phase. The outburst model results in an unabsorbed bolometric flux (0.01-100.0 keV) of $2.1_{-0.6}^{+0.2}\times10^{-9}\;erg\;cm^{-2}\;s^{-1}$ 
and a corresponding luminosity of $\sim3.3\times10^{33}\;erg\;s^{-1}$. Using eqn. \ref{eqn:luminosity}, we obtain $\dot{m}$ $\sim 1.7\times10^{16}\;g\;s^{-1}$ $\sim 2.7\times10^{-10}\;M_{\odot}\; yr^{-1}$ during the outburst. This enhanced $\dot{m}$ during outburst ($\sim 35$ times of quiescence phase) is expected for an optically thick boundary layer \citep{Patterson_1985b}. We clearly notice the signs of the blackbody emission in the soft X-rays from our modelling, which is consistent with an optically thick boundary layer. The temperature of the blackbody ($T_{\text{BL}}$) is $28_{-2}^{+1}$ eV,  along with an unabsorbed blackbody flux (0.01-100.0 keV) of $2.0_{-0.2}^{+0.8}\times10^{-9}\;erg\;cm^{-2}\;s^{-1}$ and a corresponding luminosity ($L_{\text{BL}}$) of  $\sim3.2\times10^{33}\;erg\;s^{-1}$. This agrees with the expectation that the increased luminosity in X-rays during outburst emits almost entirely ($\sim97\%$) in the very soft X-rays in the form of blackbody from the optically thick BL. Following \cite{Frank_2002}, we can relate the blackbody luminosity to the area of the emitting region as follows,

\begin{equation}
    L_{\text{BL}} = 4\pi R_{\text{WD}} H \sigma T_{\text{BL}}^4
\end{equation}
where $\sigma$ is the Stefan–Boltzmann constant, and H denotes the radial extent of the emitting region. Substituting for $L_{\text{BL}}, T_{\text{BL}}$ and $R_{\text{WD}}$, we obtain $H\sim1.0\times10^{6}\;\text{cm} \sim 2.6\times10^{-3}R_{\text{WD}}$. This result shows that the optically thick BL is very close to the WD surface, which is consistent with the picture that the disk's inner edge reaches almost near the WD surface.

\subsection{Clues from Reflection and Partial absorber: Location of X-ray emitting plasma} \label{sec:dis_reflecction}

The Compton reflection is undoubtedly confirmed during both phases from our broadband spectral analysis. During quiescence, there is a degeneracy between the intrinsic partial absorber and the Compton reflection, which influence an overlapping energy band. Considering model Q1 involving both of those two components, we obtain a reflection amplitude of $0.40_{-0.19}^{+0.47}$. 

In a simplistic geometry, the reflection from cold neutral material is $\sim1$ when a point source emitter subtends a $2\pi$ solid angle over the reflector, i.e. emitter is just above the reflector. The reflection amplitude ($\Omega/2\pi$, where $\Omega$ is the solid angle subtended by the emitter on the reflector) thus can be related to the height of the emitter ($h$) from the reflector as (see \cite{Oliveira_2019}), $h/(R_{\text{WD}})=\sqrt{1/(1-(1-\Omega/2\pi)^{2})} - 1$.
We visualise a scenario in which the hard X-rays from optically thin BL plasma get reflected by the WD surface. Assuming a point like plasma and using the obtained value of the solid angle subtended by the optically thin BL on the surface of the WD, we calculate a height, $0.25_{-0.24}^{+0.38}\;R_{\text{WD}}$, of the plasma from the WD surface. Though this quantity is very poorly constrained, the central value represents that the inner edge of the disk is quite far away from WD surface compared to the typical values obtained for other dwarf nova. For example, eclipsing dwarf nova Z Cha \citep{Nucita_2011}, HT Cas \citep{Nucita_2009, Mukai_1997}, OY Car \citep{Ramsay_2001, Wheatley_2003} shows that extent of the X-ray emission region is slightly smaller than that the WD radius or almost comparable within few per cent more than WD radius. This makes the large value of the height of the X-ray emitting boundary layer from the WD surface ($\sim25\% R_{\rm WD}$) doubtful, which deviates from the typical picture of dwarf nova. The next concern of this scenario is that the inner edge of the disk, which is considerably far away from the WD surface, cannot be contributing to the partial absorber, especially when the binary inclination angle is relatively small $\sim37\degr$. As a consequence, it becomes difficult to explain such a strong ($\sim10^{23}\; cm^{-2}$) partial absorber during the quiescence phase. Due to these ambiguities, we do not favour the scenario based on model Q1. 

For the model Q2 in quiescence, the reflection amplitude is more than 1 ($\sim1.25$) and demands multiple sites of reflection. In this scenario, the photons emitted from the X-ray emitting plasma will get Compton scattered by both the WD surface as well as the accretion disk.

For the outburst phase, the reflection amplitude is almost similar ($\sim1.31$), within statistical uncertainty, to that of the quiescence phase value and a similar reflection geometry can be envisaged. We assume a picture where the hard X-ray emitting corona is present above the optically thick BL. Since the disk's inner edge is almost near the WD surface (as supported by the calculation in Section \ref{sec:dis_xray}), the X-rays from the emitted source can be reflected by both the WD surface and the cold accretion disk. Also, our spectral analysis showed that the outburst exhibit presence of a partial absorber (column density of $3.4_{-0.9}^{+1.0}\times10^{22}\;cm^{-2}$ with a covering fraction of $0.23^{+0.02}_{-0.02}$), intrinsic to the source. During the outburst, the accretion disk remains turbulent. The partial absorber we see during this phase may originate from the outflowing accretion disk wind.

\cite{Done_1997} found that the reflection amplitude is higher ($\sim2.2$) during the outburst than the quiescence ($\sim0.7$). They conclude that in outburst, the BL extends over the WD surface, thus providing more reflection site - WD surface as well as the inner disk. In contrast, in the quiescence phase, the inner disk is truncated or optically thin, so no reflection from the disk. But we should note that this conclusion on the value of reflection amplitudes has limitations. Their spectral analysis, based on GINGA observation, already fits a simpler continuum model with reasonably good statistics even without considering reflection.

On the other hand, \cite{Ishida_2009} obtained much higher reflection amplitude in quiescence ($\sim1.7$) than in outburst ($\sim0.9$). Their explanation is quite the opposite to \cite{Done_1997}, where they describe the reflection of the emission by BL from both the accretion disk and the WD surface in quiescence. But, in the outburst, the optically thin thermal plasma is distributed as corona over an optically thick accretion disk, thereby getting reflected only from the accretion disc. But the difficulty with their measured reflection amplitudes is that either the data is not directly extended till Compton reflection hump  (in outburst, using only Suzaku XIS data extended till 10 keV), or the complications due to systematic uncertainties in the HXD/PIN background level and/or cross-calibration between XIS and HXD/PIN (in quiescence, using simultaneous XIS and HXD/PIN data in 4.2-40 keV).

A recent study \citep{Kimura_2021} of anomalous outburst event of SS Cyg in 2021 using simultaneous NICER and NuSTAR data showed a weak reflection ($\sim0.2$) amplitude during the anomalous outburst event as well as during the quiescence phase before it. They proposed a large shock height where the hard X-ray emitting corona is expanded in the vertical direction greatly, similar to a coronal siphon flow scenario, to an extent which is more than the WD radius and resulting in a weak reflection (their Figure 10). However, it would be unusual (see the results on eclipsing dwarf novae mentioned earlier) for the inner edge of the disk to be truncated so far away from the WD radius ($\sim10 R_{\rm WD}$) during quiescence, and the X-ray emitting plasma is extended till such a great distance. Note also that the maximum plasma temperature during quiescence epochs T1 and T2 derived by Kimura et al. are higher ($\sim 33$ keV) than our value. More importantly, the Fe abundance value obtained by Kimura et al. of $\sim0.10$ is significantly lower than our value ($\sim0.5$) or that of Ishida et al. ($\sim$0.37). \cite{Harrison_2015}, using infrared spectroscopy of the secondary, also found a Fe abundace of $\sim 0.3-1.0 Z_{\odot}$, even though they noted the potential degeneracy between the secondary temperature and abundance. We note that there is correlation among the reflection amplitude, the plasma upper temperature and the Fe abundance, and thereby the measurement of reflection amplitude can be impacted by the other two parameter values. In particular, while the reflection amplitude and the maximum temperature can change from one state to another, the abundance is expected to remain constant, so it is essential to reconcile the discrepant abundance values. Finally, we note the possibility that the result of Kimura et al. potentially depends on perfect cross-calibration between NICER and NuSTAR, which may be unrealistic considering the relatively young age of the NICER mission.

At this point, our study, based on the simultaneous XMM-Newton and NuSTAR data having good cross-calibration among the instruments and extended broadband coverage, securely proves the existence of the reflection. Our measured reflection amplitudes are similar and higher than 1, indicating that both the disk and the WD surface are responsible for the reflection in both phases. We agree that there are disagreements of our results with the previous studies. Although we have our reasons to believe our modelling is correct, the disagreements should be resolved in future studies.

Now, another important feature is the neutral Fe $K_{\alpha}$ line, whose parameters agree within statistical uncertainty (see Table \ref{tab:iron}) in both phases. This is expected because the strength of the neutral iron is correlated with the reflection amplitude as both the fluorescence emission and the Compton reflection originate from the cold material like the WD surface or accretion disk. The Gaussian line width of the neutral line is consistent with the instrument resolution during both the quiescence and outburst phases. We should note that the actual widths of the Fe lines are limited by the instrument resolution of the EPIC detectors ($ \Delta E \sim 128$ eV at 6.4 keV). Therefore, it is hard to distinguish from the EPIC data whether an actual narrow component of this line is contributed from the WD surface and a relatively broader component is contributed from the accretion disk. As a matter of fact, the disk contribution is unlikely to be resolved with EPIC data, given the possible Keplarian velocity near WD surface along the line of sight ($ v_{\rm k} \sin{i} \sim 3700 \rm km\,s^{-1}$, for the given WD mass and radius of SS Cyg and its binary inclination ($i$)) can at most produce a line width of $\sim 80$eV. Also, the high-resolution spectroscopic study \citep{Rana_2006} of neutral Fe line of SS Cyg in both the quiescence and outburst phase using HETG data suggest that the lines are of similar width ($\sim60$ eV). The relative contribution of WD surface and the disk to Compton reflection can be calculated based on the strength of the narrow and broad components of the neutral Fe K$_\alpha$ line. In that regard, future observations with high-resolution instruments, e.g., XRISM, has the potential to determine this via a secure deconvolution of the neutral Fe K$_\alpha$ line into a narrow and broad component.

\section{Summary}

In this work, we have investigated the dwarf nova SS Cyg during quiescence and outburst using broadband X-ray spectroscopy carried out by XMM-Newton and NuSTAR. Thanks to the good cross-calibration between the two telescopes and excellent sensitivity in broad energy bands - in soft X-rays (XMM-Newton) and hard X-rays (NuSTAR), we have distinguished spectral features and thus physical conditions in both phases, and constrained the WD mass. In summary, we found that:

\begin{itemize}

\item The spectral modelling showed that the spectra were harder during quiescence (upper temperature of the plasma emission component is $\sim22.8$ keV) than the  outburst phase ($\sim8.2$ keV).
\item The bolometric (0.01-100.0 keV) luminosity (as well as the mass accretion rate) during outburst is $\sim35$ times more than quiescence. The major contribution to flux during outburst appears from the optically thick boundary layer in the form of blackbody emission ($T_{\rm BL} \sim 28\;\rm eV$). During the outburst, the inner edge of the disc resides very close to the WD surface ($\sim0.0026R_{\rm WD}$).
\item Both the quiescence and outburst spectra showed similarly strong reflection amplitude, demanding both the disk and the WD surface to contribute to the reflection in both phases. The strengths of the neutral iron K$_{\alpha}$ lines are also similar between both phases. Our current data have limitations to resolve in the broad and narrow components of the line, which could potentially reveal the relative contribution of the disk and the WD surface to Compton reflection.
\item Partial intrinsic absorber ($\sim3\times10^{22}\; \rm cm^{-2}$ with covering fraction $\sim0.23$), possibly arising from outflowing accretion disk wind, is present during the outburst. For quiescence, our modelling exhibited a delicate statistical degeneracy between the partial absorber and the reflection component. However, we could not physically justify the presence of the partial intrinsic absorber during this phase.
\item The WD mass is consistent with $1.18_{-0.01}^{+0.02} \rm M_{\odot}$, as inferred from the X-ray broadband spectral modelling of the quiescence phase.

\end{itemize}

RLO was partially supported by the Brazilian institution CNPq (PQ-312705/2020-4). 
This research has used the data obtained from the NuSTAR telescope, operated jointly by Caltech and NASA, and the XMM-Newton telescope, operated by ESA. We thank the NuSTAR science operation team and the XMM-Newton science operation team for providing the data. We acknowledge the members at the helpdesk maintained by the High Energy Astrophysics Science Archive Research Center (HEASARC) and the XMM-Newton helpdesk team members for providing necessary support. The data used for analysis in this article are publicly available in NASA’s High Energy Astrophysics Science Archive Research Center (HEASARC) archive (\url{https://heasarc.gsfc.nasa.gov/docs/archive.html}) and XMM-Newton Science archive (\url{http://nxsa.esac.esa.int/nxsa-web/#search}). We also acknowledge the variable star observations from the AAVSO International Database contributed by observers worldwide and used in this research.

\facilities{XMM-Newton, NuSTAR}

\software{HEASoft (\cite{Heasarc_2014}, \url{https://heasarc.gsfc.nasa.gov/docs/software/heasoft/}),  
          XMM-Newton SAS (\cite{Gabriel_2004}, \url{https://www.cosmos.esa.int/web/xmm-newton/sas}), 
          }

\bibliography{draft}{}
\bibliographystyle{aasjournal}

\end{document}